\documentclass{llncs}
\usepackage{makeidx}  
\usepackage{latexsym}
\usepackage{amsmath}
\usepackage{amsfonts}
\usepackage{epsfig}
\usepackage{color}
\usepackage{graphicx}
\usepackage{boxedminipage}
\begin{document}
\pagestyle{headings}  

\newtheorem{observation}{\bf Observation}

\newcommand{\no}{\noindent}
\newcommand{\bea}{\begin{eqnarray}}
\newcommand{\eea}{\end{eqnarray}}
\newcommand{\beq}{\begin{equation}}
\newcommand{\eeq}{\end{equation}}
\newcommand{\beqs}{\begin{equation*}}
\newcommand{\eeqs}{\end{equation*}}
\newcommand{\beas}{\begin{eqnarray*}}
\newcommand{\eeas}{\end{eqnarray*}}
\newcommand{\ep}{\end{proof}}
\newcommand{\bp}{\begin{proof}}
\newcommand{\D}{\displaystyle}

\newenvironment{statement}{\begin{trivlist}
\item{\bf Statement:}}{\it \end{trivlist}}

\newenvironment{internalproof}{\begin{trivlist}
\item{\bf Proof:}}
{\hfill$\Box$\end{trivlist}}

\def\fct#1{{\mathop{\rm #1}}}   
\def\abs{\fct{abs}}             
\def\sign{\fct{sign}}           
\def\re{\fct{Re\,}}             
\def\im{\fct{Im\,}}             
\def\argmax{\mathop{\rm argmax}}
\def\argmin{\mathop{\rm argmin}}

\def\hG{{\mathbf{p}}}
\def\hg{{\mathbf{p}}}
\def\hC{\widehat{C}}
\def\hc{\widehat{C}}
\def\hl{\frac{1}{2} \log_2 }
\def\PBS0{\frac{\widehat{h}_b P_T}{M} }
\def\PG{P^g  }
\def\PB{P^b  }
\def\PGS{P^{g^*}  }
\def\PBS{P^{b^*}  }
\def\RG{R^g  }
\def\RB{R^b  }
\def\a{\alpha }
\def\DH{\Delta \! H }
\def\-{  \!-\! }
\def\+{  \!+\! }
\def\argmax{\operatornamewithlimits{arg\,max}}
\def\mbf{\mathbf}
\def\hsA{\hspace*{0.3in}}
\def\hsB{\hspace{0.5cm}}
\def\f{\frac{1}{2}}
\def\o{\overline}
\def\n{\nonumber}
\def\b{\bar}
\def\hP{\hat{P}}
\def\mb{\mbox}

\newcommand{\remove}[1]{}
\def\N{{\cal N}}
\def\R{{\cal R}}
\def\Q{{\cal Q}}
\def\p{{\mathbf{p}}}
\def\q{{\mathbf{q}}}
\def\pm{{\mathbf{p_{\min}}}}
\def\C{\widetilde C}
\def\D{\widehat D}
\def\paths{{\cal P}}
\def\NE{{Nash-routing}}

\mainmatter              

\titlerunning{\it Intl.Conf on Distributed Computing in Sensor Systems (DCOSS06)}

\title{Bottleneck Routing Games with Low Price of Anarchy}

\author{Rajgopal Kannan$^\dagger$ \and Costas Busch$^\dagger$ }


\institute{$^\dagger$Dept of Computer Science, Louisiana State University, Baton Rouge, LA 70803  \\
}

\maketitle

\begin{abstract}
We study {\em bottleneck routing games} where the
social cost is determined by the worst congestion
on any edge in the network.
In the literature, bottleneck games assume
player utility costs determined by the worst congested edge in
their paths.
However,
the Nash equilibria of such games are inefficient since the price of anarchy can
be very high and proportional to the size of the network.
In order to obtain smaller price of anarchy we
introduce {\em exponential bottleneck games}
where the utility costs of the players
are exponential functions of their congestions.
We find that exponential bottleneck games are very efficient and give
a poly-log bound on the price of anarchy: $O(\log L \cdot \log |E|)$,
where $L$ is the largest path length in the players' strategy sets and $E$
is the set of edges in the graph.
By adjusting the exponential utility costs with a logarithm
we obtain games whose player costs are almost identical
to those in regular bottleneck games,
and at the same time have the good price of anarchy of
exponential games.
\end{abstract}



\section{Introduction}
\label{section:intro}
Motivated by the selfish behavior of entities in communication
networks, we study routing games in general networks
where each packet's path is controlled independently by a selfish player.
We consider non-cooperative games with $N$ players,
where each player has a {\em pure strategy profile}
from which it selfishly selects a single
path from a source node to a destination node
such that the selected path minimizes the player's utility cost function
(such games are also known as {\em atomic} or {\em unsplittable-flow} games).
We focus on {\em bottleneck games} where the objective for the social outcome
is to minimize $C$, the maximum congestion on any edge in the network.
Typically, the congestion on an edge is a non-decreasing function
on the number of paths that use the edge;
here, we consider the congestion to be simply the number of paths that use the edge.

Bottleneck congestion games have been studied
in the literature \cite{BO07,BM06,BKV08,BKV09}
where each player's utility cost is
the worst congestion on its path edges.
In particular, player $i$ has utility cost function
$C_i = \max_{e \in p_i} C_e$
where $p_i$ is the path of the player and
$C_e$ denotes the congestion of edge $e$.
In \cite{BO07} the authors observe that
bottleneck games are important in networks for various practical reasons.
In wireless networks the maximum congested edge is related to the
lifetime of the network since the nodes adjacent to high congestion edges
transmit large number of packets which results to
higher energy utilization.
Thus, minimizing the maximum edge congestion immediately translates
to longer network lifetime.
High congestion edges also result to congestion hot-spots in the network which
may slow down the performance of the whole network.
Hot spots may also increase the vulnerability of the network
to malicious attacks which aim to to increase the congestion of edges
in the hope to bring down the network or degrade its performance.
Thus, minimizing the maximum congested edge results to hot-spot avoidance
and also to more secure networks.

Bottleneck games are also important from a theoretical point of view
since the maximum edge congestion is immediately related to the
optimal packet scheduling.
In a seminal result,
Leighton {\it et al.} \cite{LMR94}
showed that there exist packet scheduling algorithms that
can deliver the packets along their chosen paths in time very close to $C+D$,
where $D$ is the maximum chosen path length.
When $C \gg D$,
the congestion becomes the dominant factor in the packet scheduling performance.
Thus, smaller $C$ immediately implies faster packet delivery time.

A natural problem that arises concerns the effect of the players'
selfishness on the welfare of the whole network
measured with the {\em social cost} $C$.
We examine the consequence of the selfish behavior
in pure {\em Nash equilibria} which are stable states of the game in which no
player can unilaterally improve her situation.
We quantify the effect of
selfishness with the {\em price of anarchy}
($PoA$)~\cite{KP99,P01}, which expresses how much larger is the
worst social cost in a Nash equilibrium compared to the social cost
in the optimal coordinated solution.
The price of anarchy provides a measure for estimating
how closely do Nash equilibria of bottleneck routing games approximate the
optimal $C^*$ of the respective coordinated routing optimization problem.

Ideally, the price of anarchy should be small.
However, the current literature results have only provided weak bounds for bottleneck games.
In \cite{BO07} it is shown that if the edge-congestion function is bounded
by some polynomial with degree $d$ (with respect to the packets that use the edge)
then $PoA = O(|E|^d)$,
where $E$ is the set of edges in the graph.
In \cite{BM06} the authors consider the case $d=1$ and they show that $PoA = O(L + \log |V|)$,
where $L$ is the maximum path length in the players strategies and $V$ is the set of nodes.
This bound is asymptotically tight since there are game instances with $PoA = \Omega(L)$.
Note that $L \leq |E|$, and further $L$ may be significantly smaller than $|E|$.
However, $L$ can still be proportional to the size of the graph,
and thus the price of anarchy can be large.

\subsection{Contributions}

The lower bound in \cite{BM06} suggests that in order to obtain
better price of anarchy in bottleneck games (where the social cost is the bottleneck edge $C$),
we need to consider alternative player utility cost functions.
Towards this goal, we introduce {\em exponential bottleneck games}
whose social cost is the bottleneck $C$ and the player cost functions
are exponential expressions of the player congestions along their paths.
These games can be  easily converted 
to ``almost'' regular bottleneck games that preserve
the good price of anarchy of the exponential games.

In the exponential bottleneck games
the player utilities are exponential functions
on the congestion of the edges along the chosen paths.
In particular, the player utility cost function for player $i$ is:
$\C_i = \sum_{e \in p_i} 2^{C_e}.$
Note that the new utility cost is a sum of exponential
terms on the congestion of the edges in the path
(instead of the max that we described earlier).
Using the new utility cost functions
we show that exponential games have always Nash equilibria
which can be obtained by best response dynamics.
Furthermore,
for the bottleneck social cost $C$
we prove that the price of anarchy is poly-log:
\begin{equation}
PoA = O(\log L \cdot \log |E|),
\label{eqn:basic}
\end{equation}
where $L$ is the maximum path length in the players strategy set
and $E$ is the set of edges in the graph.
This price of anarchy bound is a significant improvement over the
price of anarchy from the regular
utility cost functions described earlier.

It can be shown that exponential games can be easily converted to
equivalent games with player cost $C'_i$ which are closely related to the bottleneck cost $C_i$.
In particular, 
we can obtain equivalent
games which have
similar stabilization properties while exactly preserving the price of anarchy
by taking the monotonic cost function $C'_i = \log(\C_i)$.
It holds that $C_i \leq C'_i \leq C_i + \log n$,
where $n$ is the number of nodes in the graph.
Thus, in the resulting game with utility cost $C'_i$, the player cost functions
are very close to $C_i$, and also the price of anarchy is the same as in Equation \ref{eqn:basic}.

Exponential games are interesting variations of bottleneck games
not only because they can provide good price of anarchy but also
because they represent real-life problems.
It has been shown that in wireless networks the power used by individual
nodes to transmit messages along an edge with guaranteed rate is
exponentially proportional to the flow of the edge.
Thus, exponential game equilibria represent also power game equilibria in
wireless networks, where small price of anarchy translates to small
power utilization by the nodes.
Exponential cost functions on edge congestion
have been used before in a different context for online routing optimization problems \cite{BorodinBook}[Chapter 13].
However, here we use the exponential functions for the first
time in the context of routing games.
Our technical proofs are based on the novel idea
of proving the existence and exploring properties
of {\em expansion chains} of players and edges in Nash equilibria.
This technique differs significantly from the potential
function analysis used in other literature.

\subsection{Related Work}

Koutsoupias and Papadimitriou \cite{KP99} introduced
the notion of price of anarchy
in the specific {\em parallel link networks} model
in which they provide the bound $PoA = 3/2$.
Roughgarden and Tardos \cite{roughgarden3}
provided the first result for splittable flows in general networks
in which they showed that $PoA\le 4/3$ for a player
cost which reflects to the sum of congestions of the edges of a path.
Pure equilibria with atomic flow have been studied in
\cite{BM06,CK05,libman1,STZ04} 
(our work fits into this category),
and with splittable flow in
\cite{roughgarden1,roughgarden2,roughgarden3,roughgarden5}.
Mixed equilibria with atomic flow have been studied in
\cite{czumaj1,GLMMb04,KMS02,KP99,LMMR04,MS01,P01},
and with splittable flow in
\cite{correa1,FKS02}.

Most of the work in the literature uses a cost metric
measured as the sum of congestions of all the edges of the player's path
\cite{CK05,roughgarden2,roughgarden3,roughgarden5,STZ04}.
Our work differs from these approaches since we adopt
the exponential metric for player cost.
The vast majority of the work on routing games has been performed
for parallel link networks,
with only a few exceptions on general network topologies
\cite{BM06,CK05,correa1,roughgarden1},
which we consider here.

Our work is close to \cite{BM06},
where the authors consider the player cost $C_i$ and social cost $C$.
They prove that the price of stability is 1.
They show that the price of anarchy is bounded by $O(L + \log n)$,
where $L$ is the maximum allowed path length.
They also prove that $\kappa \leq PoA \leq c(\kappa^2 + log^2 n)$,
where $\kappa$ is the size of the largest edge-simple cycle in the graph and $c$ is a constant.
That work was extended in \cite{BKV08,BKV09} to the $C+D$ routing problem.
Bottleneck congestion games have also been studied in~\cite{BO07},
where the authors consider the maximum congestion metric in general networks
with splittable and atomic flow (but without considering path lengths).
They prove the existence and non-uniqueness
of equilibria in both the splittable and atomic
flow models.
They show that finding the best Nash equilibrium that
minimizes the social cost is a NP-hard problem.
Further, they show that the price of anarchy may be unbounded for specific
edge congestion functions.

%


\section{Definitions}
\label{section:definitions}

\subsection{Path Routings}
Consider an arbitrary graph $G = (V,E)$ with nodes $V$ and edges $E$.
Let $\Pi = \{ \pi_1, \ldots, \pi_N \}$
be a set of packets such that each $\pi_i$ has
a source $u_i$ and destination $v_i$.
A {\em routing}
$\p=[p_1,p_2,\cdots,p_N]$ is a collection of paths,
where $p_i$ is a path for packet $\pi_i$ from $u_i$ to $v_i$.
We will denote by $E(p_i)$ the set of edges in path $p_i$.
Consider a particular routing $\p$.
The \emph{edge-congestion} of an edge $e$, denoted $C_e$,
is the number of paths in $\p$ that use edge $e$.
For any set of edges $A \subseteq E$,
we will denote by $C_{A} = \max_{e \in A} C_e$.
For any path $q$, the \emph{path-congestion} is $C_q = C_{E(q)}$.
For any path $p_i \in \p$,
we will also use the notation $C_i = C_{p_i}$.
The \emph{network congestion} is $C = C_{E}$,
which is the maximum edge-congestion over all edges in $E$.

We continue with definitions of exponential functions on congestion.
Consider a routing $\p$.
For any edge $e$, we will denote $\C_e = 2^{C_e}$.
For any set of edges $A \subseteq E$, we will denote $\C_{A} = \sum_{e \in A} \C_e$.
For any path $q$, we will denote $\C_q = \C_{E(q)}$.
For any path $p_i \in \p$ we will denote $\C_{i} = \C_{p_i}$.
We denote the length (number of edges) of any path $q$ as $|q|$.
Whenever necessary we will append $(\p)$ in the above definitions
to signify the dependance on routing $\p$.
For example, we will write $C(\p)$ instead of $C$.

\subsection{Routing Games}
A routing game in graph $G$ is a tuple
$\R = (G,\N,\paths)$, where
$\N$
is the set of $N$ players such
that each corresponds to one of the packets $\pi_i$
with source $u_i$ and destination $v_i$,
and $\paths$ are the strategies of the players.
In the set $\paths = \bigcup_{i \in \N} \paths_i$
the subset $\paths_i$ denotes the {\em strategy set} of player $\pi_i$
which a collection of
available paths in $G$ for player $\pi_i$ from $u_i$ to $v_i$.
Any path $p \in \paths_i$
is a {\em pure strategy} available to player $\pi_i$.
A {\em pure strategy profile} is any routing $\p=[p_1,p_2,\cdots,p_N]$,
where $p_i \in \paths_i$.
The {\em longest path length} in $\paths$
is denoted $L(\paths) =\max_{p\in \paths}|p|$.
(When the context is clear we will simply write $L$).

For game $R$ and routing $\p$,
the \emph{social cost} (or {\em global cost})
is a function of routing $\p$, and it is
denoted $SC(\p)$.
The \emph{player or local cost} is also a function on $\p$
denoted $pc_i(\p)$.
We use the standard notation
$\p_{-i}$ to refer to the collection of paths
$\{p_1,\cdots,p_{i-1},p_{i+1},\cdots,p_N\}$, and
$(p_i;\p_{-i})$ as an alternative notation for $\p$ which
emphasizes the dependence on $p_i$.
Player $\pi_i$
is \emph{locally optimal} (or {\em stable}) in routing $\p$ if
$pc_i(\p) \leq pc_i(p_i';\p_{-i})$ for
all paths $p_i'\in\paths_i$.
A greedy move by a player $\pi_i$ is any change of its path from $p_i$ to $p'_i$
which improves the player's cost,
that is, $pc_i(\p) > pc_i(p'_i;\p_{-i})$.
{\em Best response dynamics} are sequences of greedy moves by players.

A routing $\p$ is in a Nash Equilibrium
(we say $\p$ is a \emph{Nash-routing})
if every player is locally optimal.
Nash-routings quantify the notion of
a stable selfish outcome.
In the games that we study there could exist multiple Nash-routings.
A routing $\p^*$ is an optimal pure strategy profile
if it has minimum
attainable social cost: for any other pure strategy profile
$\p$, $SC(\p^*)\le SC(\p)$.

We quantify the quality of the Nash-routings with the
\emph{price of anarchy} ($PoA$)
(sometimes referred to as the coordination ratio)
and the
\emph{price of stability} ($PoS$).
Let $\bf P$ denote the set of distinct
Nash-routings, and let
$SC^*$ denote the
social cost of an optimal routing $\p^*$.
Then,
\begin{equation*}
PoA
=\sup\limits_{\p\in ~{\bf P}} \frac{SC(\p)}{SC^*},
\qquad
PoS
= \inf\limits_{\p\in ~{\bf P}} \frac{SC(\p)}{SC^*}.
\end{equation*}

\section{Exponential Bottleneck Games and their Stability}
\label{section:stability}

Let $\R = (G,\N,\paths)$
be a routing game
such that for any routing $\p$
the social cost function is $SC = C$,
and the player cost function is $pc_i = \C_i$.
We refer to such routing games as {\em exponential bottleneck games}.

We show that exponential games have always Nash-routings.
We also show that there are instances of exponential games
that have multiple Nash-routings.
The existence of Nash routings relies on finding
an appropriate potential function that provides an ordering of the routings.
Given an arbitrary initial state
a greedy move of a player can only
give a new routing with smaller order.
Thus, best response dynamics (repeated greedy moves)
converge to a routing where no player can improve further,
namely, they converge to a Nash-routing.
The potential function that we will use is:
$f(\p) = \C_{E}(\p)$.
We show that any greedy move gives a new routing with lower potential.

\begin{lemma}
\label{theorem:greedy}
If in routing $\p$ a player $\pi_i$ performs a greedy move,
then the resulting routing $\p'$ has $\C_E(\p) > \C_E(\p')$.
\end{lemma}


Since the result of the potential function cannot be smaller than zero,
Lemma \ref{theorem:greedy} implies that best response dynamics converge to Nash-routings.
Thus, we have:

\begin{theorem}
\label{theorem:stability}
Every exponential game instance $\R = (G,\N,\paths)$ has a Nash-routing.
\end{theorem}

We would like to note that there exist instances of exponential games 
that have multiple Nash-routings. Next we bound the price of anarchy 
with respect to the worst Nash-routing.

%
%
%

\section{Price of Anarchy}
\label{section:anarchy}
We bound the price of anarchy in exponential bottleneck games.  Consider
an exponential bottleneck routing game $\R = (G,\N,\paths)$.  Let $\p =
[p_1, \ldots, p_N]$ be an arbitrary Nash-routing with social cost $C$;
from Theorem \ref{theorem:stability} we know that $\p$ exists.  Let $\p^*
= [p^*_1, \ldots, p^*_N]$ represent the routing with coordinated optimal
social cost $C^*$.  We will obtain an upper bound on the price of anarchy
$PoA \leq C/C^*$.  In the analysis, we will use several parameters as
defined in the table below.
The proof relies on the notion of self-sufficient set of players.
%
\begin{center}
\begin{tabular}{|l|l|}
\hline
Param. & Definition\\
\hline
$\p, \p^*$ & Nash routing and optimal routing, respectively\\
$C, C^*$ & congestion in $\p$ and $\p^*$, respectively\\
$L$ & maximum allowed path length in players' strategy set $\paths$\\
$L^*$ & longest path length in optimal routing $\p^*$ (notice that $L^* \leq L$)\\
$\hC, 2^{\hc}$ &upper bound on congestion, upper bound on player costs in $\p$ \\
$l^*, l^*_1$ & $l^* = \log L^*$, $l^*_1 = \log (L^*-1)$ (logarithms are base 2)\\
\hline
\end{tabular}
\end{center}
%
%
\begin{definition}[Self-sufficient player set]
{\it Consider an arbitrary set of players $S$ in Nash-routing $\p$.
For each player $\pi_i \in S$
let $\q_i$ be the routing where all players in $S$ have the same paths
as in $\p$ except for player $i$ whose path is now $p^*_i$ (there are no paths in $\q_i$ 
other than for players in $S$).
We label the set of players $S$ as self-sufficient in $\p$
if for each $\pi_i \in S$ it holds $pc_i(\q_i) \geq pc_i(\p)$.
Namely, in routing $\p$ player $\pi_i$ does not switch to optimal path
$p_i^*$ only because of congestion caused by players in $S$.}
\end{definition}

Trivially, in a Nash-routing, the set of all players is self-sufficient.
If $S$ is not self-sufficient, then $\forall i \in S, p^*_i$ are called
{\it expansion edges} because additional players $S'$ must be present on them to
guarantee the Nash-routing.  We define the notion of support sets:

\begin{definition}[Support player set]
{\it If in Nash-routing $\p$ a set of players $S$ is not self-sufficient,
then there is a (support) set of players $S' \neq \emptyset$, where $S \cap S' = \emptyset$,
such that for each $\pi_i \in S$ it holds $pc_i(\q_i) \geq pc_i(\p)$,
where $\q_i$ is the routing where all players in $S \cup S'$ have the same paths
as in $\p$ except for player $i$ whose path is now $p^*_i$ (there are no other paths in $\q_i$).
Namely, in routing $\p$ player $\pi_i$ does not switch to optimal path
$p_i^*$ only because of congestion caused by players in $S \cup S'$.}
\end{definition}

Note that there could be multiple support sets for a non self-sufficient
set $S$ in Nash-routing $\p$.  If a set is not self-sufficient then there
is a support set $S'$.  The set $S \cup S'$ may not be self-sufficient
either, which implies the existence of a support set $S''$.  If $S
\cup S' \cup S''$ is not self-sufficient then there is some new support
set for it.  The process repeats until we find a self-sufficient set.
Every time we find a new support set, the previous set grows and we call
this {\em expansion}.


\subsection{Outline of Proof} 


Let $\hC = \lceil \max_i \log_2 \tilde{C}_i \rceil$. Let `game cost'
$2^{\hC}$ denote the maximum cost of a player in Nash-Routing $\p$
rounded to the nearest power of 2. Note that there can be many possible
Nash routings for a given game cost $\hC$. Furthermore the bottleneck
congestion (social cost) in Nash routing $\p$ is $\hC \geq C \geq \hC -
l^*$, by definition of the exponential player cost function.  We will
use $\hC/C^*$ to find the upper bound on the $PoA$.


Let $X$ be a (large) set of self-sufficient players.  Define an {\it
expansion chain} rooted at $X_1$ as an ordering $X_1 \rightarrow X_2
\rightarrow \ldots \rightarrow X_k$ of players in $X$ by {\it decreasing} cost levels
(where a level is a range of costs as defined below) and satisfying the
following properties: 1) $X = \bigcup_1^k X_i$ where each $X_i$ is a set
of players at the same cost level; 2) $X_i \bigcap X_j = \phi$;  3) No
prefix group of players $\bigcup_1^j X_i$ is self-sufficient for $1 \leq j <k$
and at least one player from their support sets is in some $X_t, t >j$.
(Note that individual $X_i$'s might be self-sufficient but the union,
starting from $X_1$ is not.)
For example, note that $X_1$'s support set might consist of players from
different levels in the chain;
4) Every player in $X_i, i>1$ is in the support set of some other $X_j$, 
$j\neq i$.



We label this as an expansion chain $EC$ because starting with $X_1$, the
support set of the $\bigcup_i X_i$'s is increasing by adding players
of lower cost. However as we keep expanding, eventually we will arrive
at a set of players at some lowest cost level $k$, who make the entire
chain seen so far self-sufficient.


For a given game cost $2^{\hC}$, we are interested in finding the {\it
minimum} sized graph (number of edges) that supports both a socially
optimal routing and Nash equilibrium routing with characteristics $C^*$
and $\hC$.  Since every Nash routing will have an associated expansion
chain, the equivalent goal is to find the minimum sized expansion chain.
Note that expansion chains bound the number of edges in the graph.
Each player in $EC$ performs an essential function in the Nash-routing
by property 4 and thus the size of $EC$ (number of players) in some
sense relates to the size of the graph $G$.  More specifically,
each player in $EC$ is in the support set for some other players and
occupies the expansion edges for these players. 

In our proof, we obtain a relationship between players on $EC$ and
the number of expansion edges they occupy in $G$.  We will show 
that any Nash-routing $\p$ with $\hC = \Omega(\log L^*)$ is guaranteed to
have a minimum sized $EC$ that is also very large.  Each $X_i$ in this
minimum $EC$ must have a support set of players with costs close to it.
The number of stages in this $EC$ grows with $l^*$, however the support
set and expansion edges for each subsequent $X_i$ grows exponentially.
By finding the minimum sized $EC$ for a given $\hC$, we then find the
smallest graph $G$ (with an exponentially large number of edges) with
the given Price of Anarchy.  Equivalently, for a graph of given size,
we can then compute the upper bound on the $PoA$ for any Nash routing.

We define our cost stages (cost levels) for expansion chains and player
types in the following manner: let $S^{(i)}$ denote the set of players
in stage $i$, $1 \leq i \leq \hc$ with player costs in range $[2^{\hC \-
i} \+ 2, 2^{\hC \-i \+ 1}]$.  In stage $i$, let $A^{(i)}$ denote the set
of all players occupying exactly one edge of congestion $\hc \- i \+
1$,  let $B^{(i)}$ denote the set of all players whose {\it maximum}
edge congestion $C'$ satisfies $\hc \- i \geq C' >  \hc \- i \- l^* \-
1$ and finally let $D^{(i)} = S^{(i)} \- A^{(i)} \-  B^{(i)} $.  $1$
is the highest stage and has at least one player of type $A$, $B$ or $D$
by definition of $\hC$. Lower stages could be empty of players.

\subsection{Price of Anarchy Bound for $C^*=1$}

For ease of exposition, assume $C^*=1$, i.e every
player in the coordinated socially optimal network (we will use the term
network or game interchangeably with the term routing) has a unique
optimal path to its destination of length at most $L^*$.  The general
$C^*$ case is proved later.  


To find minimum sized expansion chains, we first need to determine if
expansion chains of size $> 1$ exist for a given value of $\hC$. Related
to this, we also need to know how large is the set of these players.
We first prove a sufficient condition on $\hC$ for expansion chains
to exist.  Subsequently, we will derive the specific minimum sized
expansion chain and its size.

\begin{lemma}
{\it
Any non-empty player set $X^{(i)} \subseteq \{ A^{(i)} \bigcup B^{(i)} \bigcup D^{(i)}
\}$ is not self-sufficient, where $1 \leq i \leq \hC \- l_1^* \- 11$.  
\label{l^*lemma} 
}
\end{lemma}
(Please see appendix for the proofs). This leads to

\begin{theorem}
{\it
Any subset of players $S \subseteq \{ S^{(1)} \bigcup S^{(2)} \bigcup
\ldots \bigcup S^{(k)} \}$ are not self-sufficient, where $k = \hC \-
l_1^* \- 11$. Equivalently there cannot exist any expansion chains
consisting only of players from the first $k$ stages.
\label{supportsettheorem}
}
\end{theorem}

Since we are guaranteed the existence of at least one player from
$S^{(1)}$ and there exists a Nash-routing, there must be a self-sufficient
set of players including this player. By the above theorem, the expansion
chain for this set and rooted at stage 1 cannot terminate before stage
$k$. Identifying this particular minimum size expansion chain allows us
to count the minimum number of edges in $G$ and hence an upper bound on
the $PoA$.

We now want to find the minimum number of edges required to support
the Nash routing with game cost $2^{\hC}$. This corresponds to finding
the smallest expansion chain rooted at stage $1$.  By our definition,
an expansion chain consists of new players occupying the optimal path
edges of players on the previous levels.  It would seem that chains
should consist of type $B$ players since they occupy multiple edges and
thus fewer players are required.  However as the lemma below shows it
is players of type $A$ that minimize the expansion edges.




Consider an arbitrary player $\pi$ of type $B$ in $\hG$ occupying
edges $E= \{e_1,e_2, \ldots, e_k\}$  of non-increasing congestion
$c_1 \geq c_2 \geq \ldots c_k$ that are optimal edges (expansion
edges) of other players, where we assume maximum congestion $c_1
\geq 2$.  We want to answer the following question: Is there an
alternate equilibrium/game containing player(s) with the same total
equilibrium cost as $\pi$, but requiring fewer edges to support
this equilibrium cost.  Note that when comparing these two games,
the actual routing paths (i.e source-destinations) do not have to
be the same. All we need to show is the existence of an alternate
game (even with different source-destination pairs for the players)
that has the same equilibrium cost.

In particular, consider an alternate game $\p'$ in which $\pi$ is
replaced by a set $P=\{\pi_1, \pi_2, \ldots, \pi_k\}$ of type
$A$ players occupying single edges of congestion $c_1, c_2, \ldots,
c_k$, where $\pi$ and the set $P$ are also in equilibrium in
their respective games. The equilibrium cost of $\pi$ and set $P$
is the same ($\sum_{j=1}^k 2^{c_j}$) as they are occupying edges
of the same congestion.  Since both $\pi$ in game $\hG$ and the set
of players $P$ are in equilibrium and occupying expansion edges of
other players in their respective games, $C^*=1$ implies they must
have their own expansion edges in their respective games.  Suppose
we can show that the number of expansion edges required by the $k$
players in $P$ is at most those required by the single player of
type $B$.  Since $\pi$ is an arbitrary type $B$ player, this argument
applied recursively implies that all expansion edges in the game
$\hG$ should be occupied by type $A$ players to minimize the total
number of expansion edges.  Thus we will have shown that any
equilibrium with cost $\hC$ can be supported with fewer total players
if they are of type $A$ than if they are of type $B$.  Let $\pi^*$
and $P^*$ denote the expansion edges of $\pi$ and the set $P$
respectively.

\begin{lemma}
{\it
$|P^*| \leq |\pi^*|$ for arbitrary players $\pi$ and set $P$ with
the same equilibrium cost.
\label{typeB=typeA}
}
\end{lemma}

As a consequence of lemma~\ref{typeB=typeA},  we have
\begin{lemma}
{\it For $\hC > l^* \+ 11$, the expansion chain rooted in stage 1 and
occupying the minimum number of edges consists only of players of
type $A$ (other than the root).
}
\label{onlyAlemma}
\end{lemma}
%



Next we derive the size of the smallest network required to support
an equilibrium congestion of $\hC$. Without loss of generality, we
assume there exists at least one type $A$ player in stage 1, i.e a
single edge of congestion $\hC$ and derive the minimum chain rooted
at $A^{(1)}$.  From lemma~\ref{onlyAlemma}, there exists an expansion
chain rooted at $A^{(1)}$ with only type $A$ players. Among all
such expansion chains, the one with the minimum number of players
(equivalently edges, since each type $A$ player occupies a single
edge) is defined below.


\begin{theorem}
{\it
$EC_{min}$, the expansion chain with minimum number of edges that
supports a self-sufficient equilibrium rooted at $A^{(1)}$ is defined
by
$EC_{min} :
A^{(1)} \rightarrow
A^{(l^*+2)} \rightarrow
A^{(2l^*+3)} \rightarrow
A^{(3l^*+4)} \rightarrow
\ldots
\rightarrow A^{(\hC-1)}$.
Every player in $EC_{min}$ has an optimal path whose length is the
maximum allowed $L^*$. The depth of chain $EC_{min}$ is $O(\hC/l^*)$.
}
\label{minEC}
\end{theorem}

$EC_{min}$ defined in Theorem~\ref{minEC} is also the minimum sized
chain when the root players are from $B^{(1)}$ or $D^{(1)}$ although
the number of edges required in the supporting graph is slightly
different as we see later. In these cases, all stages (other than
the root) in the minimum expansion chain consist of type $A$ players
by lemma~\ref{onlyAlemma} and the proof of Theorem~\ref{minEC} is
immediately applicable in choosing the specific indices of the
expansion stages required to support the equilibrium).  As 
shown later, however, the $PoA$ is maximized when the chain is rooted at
$A^{(1)}$.

\begin{theorem}
{\it When $C^*=1$, the upper bound $\kappa$ on the Price of Anarchy
$PoA$ of Nash-routing $\hG$ is given by the minimum of
1) $ \kappa =  O(\log L^*) $ or 2)
$\kappa \big( \log (\kappa L^*) \big) \leq \log L^* \cdot \log |E|$.
\label{PoATheorem}
}
\end{theorem}

Can we get a larger upper bound on the $POA$ if the expansion chain
is rooted at $B^{(1)}/ D^{(1)}$ instead of $A^{(1)}$? To examine
this, let $\hC - q$ be the largest congestion in $\hG$, $q>0$. We
need $2^q$ such edges in order to satisfy the maximum player cost
of $2^{\hC}$. All these edges can be used as expansion edges for
other players.  From the analysis in Theorem~\ref{PoATheorem}, we
note that expansion between stages occurs at a factorial rate. Thus
using these $2^q$ edges as high up in the chain as possible (thereby
reducing the need for new expansion edges) will minimize the expansion
rate. The best choice for $q$ then is $l^*$. In this case, we have
a single player $\pi_m$ in equilibrium in $\hG$, occupying $L^*$
edges of congestion $\hC - l^*$. These $L^*$ edges are also the
optimal edges of $\pi_m$, i.e its equilibrium and optimal paths are
identical. Hence the first stage of expansion in this chain is for
the $L^* (\hC -l^* -1)$ players on the $L^*$ edges of $\pi_m$.  From
this point on the minimum sized chain for this graph is identical
to the minimum sized chain $EC_{min}$ defined above.  The total
number of edges in this chain can be computed in a manner similar
to above. While the number of edges is smaller than $EC_{min}$, it
can be shown that the $PoA$ is also smaller $\hC -l^*$.  Hence the
upper bound on the $PoA$ is obtained using an expansion chain rooted
at $A^{(1)}$.

\subsection{Price of Anarchy Bound for $C^*>1$}
So far we have assumed the optimal bottleneck
congestion $C^*=1$ in our derivations.  We now show that increasing $C^*$
decreases the $PoA$ and hence the previous derivation is the upper bound.
We first evaluate the impact of $C^* = M > 1$ on expansion chains.
Having $C^* >1$ implies that more players can share expansion edges and
thus the rate of expansion as well as the depth of an expansion chain
(if it exists) should decrease.  We first show that expansion chains
exist even for arbitrary $C^*=M$.

\begin{lemma}
{\it
Any subset of players $S \subseteq \{ S^{(1)} \bigcup S^{(2)} \bigcup
\ldots \bigcup S^{(k)} \}$ are not self-sufficient, where $k: \hC - k >
8M + l_1^* +2$.  Equivalently there cannot exist any expansion chains
consisting only of players from the first $k$ stages.
\label{C^*lemma}
}
\end{lemma}


Similarly Lemmas~\ref{typeB=typeA} and ~\ref{onlyAlemma} can be
suitably modified and the minimum sized chain in this case has
the same structure as defined in Theorem~\ref{minEC}. Analogous to
the $C^*=1$ case, the maximum $PoA$ occurs when $EC_{min}$ is rooted
at $A^{(1)}$. We calculate this $PoA$ with $C^*=M$, below.
\begin{theorem}
{\it When $C^*=M$, the upper bound $\kappa$ on the Price of Anarchy
$PoA$ of game $\hG$ is given by the minimum of
1) $ \kappa =  O(\frac{\log L^*}{M})$ or
2) $\kappa( \log (L^* \kappa) ) \leq \frac{l^* \log |E|}{M}$
}
\label{PoAC^*Theorem}
\end{theorem}

\section{Conclusions}
\label{section:conclusions}
We show by carefully selecting appropriate player
cost functions that the price of anarchy of bottleneck routing
games is poly-log with respect to
the size of the game parameters: $O(\log L \cdot \log |E|)$.
%
%
A natural question that arises is what is the impact of polynomial cost functions
to the price of anarchy.
Polynomial cost functions with low degree give high price of anarchy.
Consider the game instance in the figure 
where the player cost is $pc_i = \sum_{e \in p_i} C_e$
which is a linear function on the congestion of the edges
on the player's path.
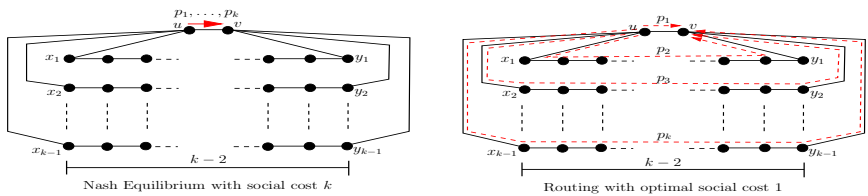
\begin{center}
\resizebox{4.5in}{1in}{\begin{picture}(0,0)%
\includegraphics{sumC.pstex}%
\end{picture}%
\setlength{\unitlength}{4144sp}%
\begingroup\makeatletter\ifx\SetFigFont\undefined%
\gdef\SetFigFont#1#2#3#4#5{%
  \reset@font\fontsize{#1}{#2pt}%
  \fontfamily{#3}\fontseries{#4}\fontshape{#5}%
  \selectfont}%
\fi\endgroup%
\begin{picture}(10092,3099)(270,-2696)
\put(905,-502){\makebox(0,0)[rb]{\smash{{\SetFigFont{12}{14.4}{\rmdefault}{\mddefault}{\updefault}{\color[rgb]{0,0,0}$x_1$}%
}}}}
\put(4363,-502){\makebox(0,0)[lb]{\smash{{\SetFigFont{12}{14.4}{\rmdefault}{\mddefault}{\updefault}{\color[rgb]{0,0,0}$y_1$}%
}}}}
\put(2331, 27){\makebox(0,0)[rb]{\smash{{\SetFigFont{12}{14.4}{\rmdefault}{\mddefault}{\updefault}{\color[rgb]{0,0,0}$u$}%
}}}}
\put(2937, 36){\makebox(0,0)[lb]{\smash{{\SetFigFont{12}{14.4}{\rmdefault}{\mddefault}{\updefault}{\color[rgb]{0,0,0}$v$}%
}}}}
\put(923,-2005){\makebox(0,0)[rb]{\smash{{\SetFigFont{12}{14.4}{\rmdefault}{\mddefault}{\updefault}{\color[rgb]{0,0,0}$x_{k-1}$}%
}}}}
\put(4338,-1996){\makebox(0,0)[lb]{\smash{{\SetFigFont{12}{14.4}{\rmdefault}{\mddefault}{\updefault}{\color[rgb]{0,0,0}$y_{k-1}$}%
}}}}
\put(923,-1057){\makebox(0,0)[rb]{\smash{{\SetFigFont{12}{14.4}{\rmdefault}{\mddefault}{\updefault}{\color[rgb]{0,0,0}$x_2$}%
}}}}
\put(4355,-1048){\makebox(0,0)[lb]{\smash{{\SetFigFont{12}{14.4}{\rmdefault}{\mddefault}{\updefault}{\color[rgb]{0,0,0}$y_2$}%
}}}}
\put(2622,232){\makebox(0,0)[b]{\smash{{\SetFigFont{12}{14.4}{\rmdefault}{\mddefault}{\updefault}{\color[rgb]{0,0,0}$p_1, \ldots, p_k$}%
}}}}
\put(2639,-2175){\makebox(0,0)[b]{\smash{{\SetFigFont{12}{14.4}{\rmdefault}{\mddefault}{\updefault}{\color[rgb]{0,0,0}$k-2$}%
}}}}
\put(2630,-2585){\makebox(0,0)[b]{\smash{{\SetFigFont{12}{14.4}{\rmdefault}{\mddefault}{\updefault}{\color[rgb]{0,0,0}Nash Equilibrium with social cost $k$}%
}}}}
\put(6251,-531){\makebox(0,0)[rb]{\smash{{\SetFigFont{12}{14.4}{\rmdefault}{\mddefault}{\updefault}{\color[rgb]{0,0,0}$x_1$}%
}}}}
\put(9709,-531){\makebox(0,0)[lb]{\smash{{\SetFigFont{12}{14.4}{\rmdefault}{\mddefault}{\updefault}{\color[rgb]{0,0,0}$y_1$}%
}}}}
\put(7677, -2){\makebox(0,0)[rb]{\smash{{\SetFigFont{12}{14.4}{\rmdefault}{\mddefault}{\updefault}{\color[rgb]{0,0,0}$u$}%
}}}}
\put(8283,  7){\makebox(0,0)[lb]{\smash{{\SetFigFont{12}{14.4}{\rmdefault}{\mddefault}{\updefault}{\color[rgb]{0,0,0}$v$}%
}}}}
\put(6269,-2034){\makebox(0,0)[rb]{\smash{{\SetFigFont{12}{14.4}{\rmdefault}{\mddefault}{\updefault}{\color[rgb]{0,0,0}$x_{k-1}$}%
}}}}
\put(9684,-2025){\makebox(0,0)[lb]{\smash{{\SetFigFont{12}{14.4}{\rmdefault}{\mddefault}{\updefault}{\color[rgb]{0,0,0}$y_{k-1}$}%
}}}}
\put(6269,-1086){\makebox(0,0)[rb]{\smash{{\SetFigFont{12}{14.4}{\rmdefault}{\mddefault}{\updefault}{\color[rgb]{0,0,0}$x_2$}%
}}}}
\put(9701,-1077){\makebox(0,0)[lb]{\smash{{\SetFigFont{12}{14.4}{\rmdefault}{\mddefault}{\updefault}{\color[rgb]{0,0,0}$y_2$}%
}}}}
\put(7985,-373){\makebox(0,0)[b]{\smash{{\SetFigFont{12}{14.4}{\rmdefault}{\mddefault}{\updefault}{\color[rgb]{0,0,0}$p_2$}%
}}}}
\put(7985,139){\makebox(0,0)[b]{\smash{{\SetFigFont{12}{14.4}{\rmdefault}{\mddefault}{\updefault}{\color[rgb]{0,0,0}$p_1$}%
}}}}
\put(7994,-1756){\makebox(0,0)[b]{\smash{{\SetFigFont{12}{14.4}{\rmdefault}{\mddefault}{\updefault}{\color[rgb]{0,0,0}$p_k$}%
}}}}
\put(7985,-834){\makebox(0,0)[b]{\smash{{\SetFigFont{12}{14.4}{\rmdefault}{\mddefault}{\updefault}{\color[rgb]{0,0,0}$p_3$}%
}}}}
\put(7977,-2217){\makebox(0,0)[b]{\smash{{\SetFigFont{12}{14.4}{\rmdefault}{\mddefault}{\updefault}{\color[rgb]{0,0,0}$k-2$}%
}}}}
\put(7977,-2627){\makebox(0,0)[b]{\smash{{\SetFigFont{12}{14.4}{\rmdefault}{\mddefault}{\updefault}{\color[rgb]{0,0,0}Routing with optimal social cost 1}%
}}}}
\end{picture}%
}
\end{center}
%
In this game there $k$ players $\pi_1, \ldots, \pi_k$
where all the players have source $u$ and destination $v$
which are connected by edge $e$.
The graph consists of $k-1$ edge-disjoint paths from $u$ to $v$
each of length $k$.
There is a Nash equilibrium, depicted in the left figure 
where every player chooses to use a path of length 1 on edge $e$.
This is an equilibrium because the cost of each player is $k$,
while the cost of every alternative path is also $k$.
Since the congestion of edge $e$ is $k$ the social cost is $k$.
The optimal solution for the same routing problem is depicted in
the right of the figure.
where every player uses a edge-disjoint path
and thus the maximum congestion on any edge is 1.
Therefore, the price of anarchy is at least $k$.
Since we can choose $k = \Theta(\sqrt n)$, where $n$ is the number of nodes
in the graph,
the price of anarchy is $\Omega(\sqrt n)$.
%
%
%
{
\bibliographystyle{IEEEtran}
\bibliography{routing}

\begin{thebibliography}{10}
\providecommand{\url}[1]{#1}
\csname url@samestyle\endcsname
\providecommand{\newblock}{\relax}
\providecommand{\bibinfo}[2]{#2}
\providecommand{\BIBentrySTDinterwordspacing}{\spaceskip=0pt\relax}
\providecommand{\BIBentryALTinterwordstretchfactor}{4}
\providecommand{\BIBentryALTinterwordspacing}{\spaceskip=\fontdimen2\font plus
\BIBentryALTinterwordstretchfactor\fontdimen3\font minus
  \fontdimen4\font\relax}
\providecommand{\BIBforeignlanguage}[2]{{%
\expandafter\ifx\csname l@#1\endcsname\relax
\typeout{** WARNING: IEEEtran.bst: No hyphenation pattern has been}%
\typeout{** loaded for the language `#1'. Using the pattern for}%
\typeout{** the default language instead.}%
\else
\language=\csname l@#1\endcsname
\fi
#2}}
\providecommand{\BIBdecl}{\relax}
\BIBdecl

\bibitem{BO07}
R.~Banner and A.~Orda, ``Bottleneck routing games in communication networks,''
  \emph{IEEE Journal on Selected Areas in Communications}, vol.~25, no.~6, pp.
  1173--1179, 2007, also appears in INFOCOM'06.

\bibitem{BM06}
C.~Busch and M.~Magdon-Ismail, ``Atomic routing games on maximum congestion,''
  \emph{Theoretical Computer Science}, vol. 410, no.~36, pp. 3337--3347, August
  2009.

\bibitem{BKV08}
C.~Busch, R.~Kannan, and A.~V. Vasilakos, ``Quality of routing congestion games
  in wireless sensor networks,'' in \emph{Proc. 4th International Wireless
  Internet Conference (WICON)}, Maui, Hawaii, November 2008.

\bibitem{BKV09}
------, ``Approximating congestion + dilation in networks via 'quality of
  routing' games,'' 2009, submitted to {IEEE/ACM} Transactions on Networking.

\bibitem{LMR94}
F.~T. Leighton, B.~M. Maggs, and S.~B. Rao, ``Packet routing and job-scheduling
  in ${O}(congestion+dilation)$ steps,'' \emph{Combinatorica}, vol.~14, pp.
  167--186, 1994.

\bibitem{KP99}
E.~Koutsoupias and C.~Papadimitriou, ``Worst-case equilibria,'' in
  \emph{Proceedings of the 16th Annual Symposium on Theoretical Aspects of
  Computer Science (STACS)}, ser. LNCS, vol. 1563.\hskip 1em plus 0.5em minus
  0.4em\relax Trier, Germany: Springer-Verlag, March 1999, pp. 404--413.

\bibitem{P01}
C.~Papadimitriou, ``Algorithms, games, and the {Internet},'' in
  \emph{Proceedings of the 33rd Annual {ACM} Symposium on Theory of Computing
  (STOC)}, {ACM}, Ed., Hersonissos, Crete, Greece, July 2001, pp. 749--753.

\bibitem{BorodinBook}
A.~Borodin and R.~El-Yaniv, \emph{Online computation and competitive
  analysis}.\hskip 1em plus 0.5em minus 0.4em\relax New York, NY, USA:
  Cambridge University Press, 1998.

\bibitem{roughgarden3}
T.~Roughgarden and \'{E}va Tardos, ``How bad is selfish routing,''
  \emph{Journal of the ACM}, vol.~49, no.~2, pp. 236--259, March 2002.

\bibitem{CK05}
G.~Christodoulou and E.~Koutsoupias, ``The price of anarchy of finite
  congestion games,'' in \emph{Proceedings of the 37th Annual {ACM} Symposium
  on Theory of Computing (STOC)}.\hskip 1em plus 0.5em minus 0.4em\relax
  Baltimore, {MD}, {USA}: ACM, May 2005, pp. 67--73.

\bibitem{libman1}
L.~Libman and A.~Orda, ``Atomic resource sharing in noncooperative networks,''
  \emph{Telecomunication Systems}, vol.~17, no.~4, pp. 385--409, 2001.

\bibitem{STZ04}
S.~Suri, C.~D. Toth, and Y.~Zhou, ``Selfish load balancing and atomic
  congestion games,'' \emph{Algorithmica}, vol.~47, no.~1, pp. 79--96, Jan.
  2007.

\bibitem{roughgarden1}
T.~Roughgarden, ``The maximum latency of selfish routing,'' in
  \emph{Proceedings of the Fifteenth Annual {ACM}-{SIAM} Symposium on Discrete
  Algorithms (SODA)}, New Orleans, Louisiana, (USA), January 2004, pp.
  980--981.

\bibitem{roughgarden2}
------, ``Selfish routing with atomic players,'' in \emph{Proc. 16th Symp. on
  Discrete Algorithms (SODA)}.\hskip 1em plus 0.5em minus 0.4em\relax ACM/SIAM,
  2005, pp. 1184--1185.

\bibitem{roughgarden5}
T.~Roughgarden and \'{E}va Tardos, ``Bounding the inefficiency of equilibria in
  nonatomic congestion games,'' \emph{Games and Economic Behavior}, vol.~47,
  no.~2, pp. 389--403, 2004.

\bibitem{czumaj1}
Czumaj and Vocking, ``Tight bounds for worst-case equilibria,'' in \emph{ACM
  Transactions on Algorithms (TALG)}.\hskip 1em plus 0.5em minus 0.4em\relax
  ACM, 2007, vol.~3.

\bibitem{GLMMb04}
M.~Gairing, T.~L{\"u}cking, M.~Mavronicolas, and B.~Monien, ``Computing {Nash}
  equilibria for scheduling on restricted parallel links,'' in
  \emph{Proceedings of the 36th Annual {ACM} Symposium on the Theory of
  Computing (STOC)}, Chicago, Illinois, {USA}, June 2004, pp. 613--622.

\bibitem{KMS02}
E.~Koutsoupias, M.~Mavronicolas, and P.~G. Spirakis, ``Approximate equilibria
  and ball fusion,'' \emph{Theory Comput. Syst.}, vol.~36, no.~6, pp. 683--693,
  2003.

\bibitem{LMMR04}
T.~L{\"u}cking, M.~Mavronicolas, B.~Monien, and M.~Rode, ``A new model for
  selfish routing,'' \emph{Theoretical Computer Science}, vol. 406, no.~3, pp.
  187--206, 2008.

\bibitem{MS01}
Mavronicolas and Spirakis, ``The price of selfish routing,''
  \emph{Algorithmica}, vol.~48, 2007.

\bibitem{correa1}
J.~R. Correa, A.~S. Schulz, and N.~E.~S. Moses, ``Computational complexity,
  fairness, and the price of anarchy of the maximum latency problem,'' in
  \emph{Proc. Integer Programming and Combinatorial Optimization, 10th
  International {IPCO} Conference}, ser. Lecture Notes in Computer Science,
  vol. 3064.\hskip 1em plus 0.5em minus 0.4em\relax New York, {NY}, {USA}:
  Springer, June 2004, pp. 59--73.

\bibitem{FKS02}
D.~Fotakis, S.~C. Kontogiannis, and P.~G. Spirakis, ``Selfish unsplittable
  flows,'' \emph{Theoretical Computer Science}, vol. 348, no. 2-3, pp.
  226--239, 2005.

\end{thebibliography}
}


%

\end{document}